\documentstyle[aps,prl,multicol,epsf]{revtex}
\def\({\begin{equation}}
\def\){\end{equation}}
\begin{document}                
\title{Phase Fluctuations and Non-Equilibrium Josephson Effect}
\author{B. N. Narozhny and I. L. Aleiner}
\address{Department of Physics and Astronomy,
SUNY Stony Brook, Stony Brook, NY 11794}

\maketitle
\begin{abstract}
We consider a diffusive S-N-S junction with electrons in the normal
layer driven out of equilibrium by external bias.  We show that, the
non-equilibrium fluctuations of the electron density in the normal
layer cause the fluctuations of the phase of the order parameter in the
S-layers.  As a result, the magnitude of the Josephson
current in the non-equilibrium junction is significantly supressed
relative to its mean field value.

\end{abstract}
\pacs{PACS numbers: 74.40.+k, 74.50.+r, 73.40Gk, 73.50.-h}

\begin{multicols}{2}

When an electric current flows through a metallic sample, electrons in the 
metal can no longer be considered as a system in equilibrium. In particular
the electronic distribution function $f(\epsilon)$ differs considerably from
the equilibrium Fermi distribution. The low temperature two-step distribution
function was recently observed in tunneling experiments \cite{exp1}. 

The ability to change the electronic distribution by simply applying voltage
to a metallic system opens a new possibility to control the supercurrent in a
S-N-S junction \cite{exp2}. The idea is based on the description of 
the supercurrent flow in terms of the electronic states and their occupation
probabilities $f(\epsilon)$. As a function of the phase difference of the two 
superconducting layers in a S-N-S junction $\theta=\theta_1-\theta_2$ the 
supercurrent density can be written as \cite{been}

\begin{equation}
J(\theta) = \int d\epsilon f(\epsilon) j(\epsilon, \theta),
\label{jc}
\end{equation}

\noindent
where $j(\epsilon, \theta)$ is the contribution of the states with energies
between $\epsilon$ and $\epsilon +d\epsilon$. Changing the electronic 
distribution, one changes the probabilities $f(\epsilon)$ and thus can change 
the magnitude \cite{exp2} and even the direction \cite{exp3} of the 
supercurrent (in other words, making a $\pi$-junction). Some particular cases
of $j(\epsilon, \theta)$ were considered theoretically \cite{zaik}.

It is important to realize that Eq.~(\ref{jc}) was obtained within the mean 
field approximation where the phase $\theta$ is fixed. Quantum fluctuations
of the phase of the order parameter do not induce dramatic change in the 
amplitude of the Josephson energy, if the conductance of the normal layer
is large. The purpose of this Letter is to show that it may not be the case in
the non-equilibrium situation.

We show that the non-equilibrium 
fluctuations in the normal layer cause the fluctuations of the phase of the
order parameter in both S-layers, thus affecting the Josephson current in the
junction [the term $j(\epsilon, \theta)$ in Eq.~(\ref{jc})]. In particular,
the critical current (which is proportional to the Josephson energy $E_J$) 
becomes strongly suppressed by the non-equilibrium. This effect accompanies 
the change of the magnitude and even of the
direction of the supercurrent, which is due to the
term $f(\epsilon)$ in Eq.~(\ref{jc}). Thus the critical current in the 
$\pi$-junction is smaller than the mean field value, as was observed in the
experiment \cite{exp3}. Here, we will present the phenomenological derivation
which yields the same results as a calculation based on the Keldysh technique
(for similar consideration of S-N junction see Ref.~\onlinecite{elsewhere}).

To describe the non-equilibrium fluctuations we first identify the collective 
modes in the junction. For simplicity we will consider an S-N-S sandwich, 
where each layer can be considered as a 2D film. The effects of the finite 
thickness of the layers will be discussed later in the Letter.

Consider a superconducting film at zero temperature. All of the excitations 
with the energy smaller than the superconducting gap $\Delta$ are associated 
with the phase $\theta$ of the order parameter \cite{Tinkham}. In the isolated 
S-layer, the longitudinal phase fluctuations correspond to the usual 2D
plasmon with dispersion $\omega\simeq \sqrt{Q}$.
However, when a layer of normal metal is present, the 
collective mode with the linear dispersion relation appears 
\cite{elsewhere}. 
In the S-N-S junction there two such modes corresponding to each S-layer. 

The time evolution of phase $\theta$ is governed by hydrodynamic equations, 
which in the absence of external magnetic fields can be written as
\cite{Tinkham}

\begin{mathletters}
\begin{eqnarray}
&&
\dot n_s^{(i)} + \frac{1}{2e}{\bf{\vec\nabla}}\cdot\vec{j}_s^{(i)} = 0,
\label{nep}\\
&&
\nonumber\\
&&
\vec{j}_s^{(i)} = -e \pi\hbar D_s^{(i)} \nu_s^{(i)} \Delta 
{\bf\vec{\nabla}} \theta^{(i)},
\label{cur}\\
&&
\nonumber\\
&&
\hbar\dot \theta^{(i)} = 2\left(e\varphi +  
{{n_s^{(i)}}\over{\nu_s^{(i)}}}\right),
\label{jos}
\end{eqnarray}

\noindent
where the superscript $i=1,2$ labels the layer, $n_s^{(i)}$ is the 
perturbation of the carrier density in S-layers,
$\vec{j}_s^{(i)}$ is the supercurrent, and $\varphi$ is the electric
potential. We 
wrote the London equation (\ref{cur}) for a dirty superconductor and 
expressed the superfluid density through the diffusion coefficient 
$D_s^{(i)}$ in the normal state of the superconductor and  the 
thermodynamic density of states per unit area in the superconductor 
$\nu_s^{(i)}$. Also 
in Eq.~(\ref{leq}) we neglected the terms which describe the Josephson
current. We will discuss this point later.

The electron density is normal metal is governed by the continuity equation 
and the Ohm's law  
\begin{equation}
e \dot n_m + {\bf{\vec\nabla}}\cdot\vec{j}_m = 0;
\quad
\vec{j}_m = - D_m{\bf{\vec\nabla}}\left( e^2 \nu_m \varphi + e n_m\right),
\label{curm}
\label{eq:4}
\end{equation}
where $n_m$ is the carrier density,
$\vec{j}_m$ is the current and $D_m$ is the diffusion coefficient 
in the N-layer.

Fluctuations of the densities in three layers are coupled through the
Coulomb potential
\begin{equation}
\varphi = \int dr'V(r-r')
\left[ n_s^{(1)}(r')+n_s^{(2)}(r')+n_m(r') \right];
\label{eq:5}
\end{equation}
\begin{equation}
V= \frac{e^2}{r}.
\label{epn}
\end{equation}
\label{leq}
\end{mathletters}

\noindent
So far we have neglected the thickness of the metallic layer wherefore only 
the sum of the electron densities in two S-layers is coupled to the density of
electrons in the metal. Therefore in a strictly two-dimensional model of
the S-N-S junction with two identical S-layers the Josephson current (which
depends on the phase {it difference}) is not affected by the density 
fluctuations in the metal. To couple the fluctuations in the normal metal 
to the phase difference one needs to introduce some asymmetry into the 
model either by having two different S-layers or by taking into account the
non-zero thickness of the metallic layer. Here we chose the former 
(see Eq.~(\ref{cur})). The final results do not depend on the asymmetry 
explicitly, and thus are independent of this choice. 

The requirement of the consistency of Eqs.~(\ref{leq}) gives two
acoustic branches of the collective mode corresponding to the sum and
difference of the electron densities of the S-layers with dispersion
relations $\omega_{1,2}(Q) =\omega_2' - i\omega_{1,2}''$. The lifetime
of both modes is finite. These modes are similar to the Schmid-Sch\"on
mode in the vicinity of the critical temperature \cite{schm}.  The
only difference is that the normal excitations are not thermally
activated in the superconductor itself but rather exist in the normal
metallic layer close to the superconductor, however, it does not
change the charge dynamics.  In the latter (odd) mode finite lifetime
appears only due to the asymmetry of the sandwich $\delta D =
[D_s^{(1)} - D_s^{(2)}]/2$

\begin{equation}
\omega_1'=Q\sqrt
\frac{\pi\Delta D_s}{\hbar};
\; \; 
\omega_1''= {\pi\over{2}}\left({{\nu_m(\delta D)^2}\over{\nu_sD_m D_s}}\right)
{\Delta\over{\hbar}},
\label{fm}
\end{equation}

\noindent
where $\nu_s=\nu_s^{(1)}+\nu_s^{(2)}$, and $G_{s,m}$ denote dimensionless 
conductances of the superconducting (in the normal state) and normal
layers respectively: 
$G_{s,m} =2\pi\hbar \sigma_{s,m} /e^2 = 2\pi\hbar\nu_{s,m} D_{s,m}$.
The conductances are measured in units of $e^2/{2\pi\hbar}=1/(25.8 K\Omega)$.
The even mode is similar to the ``phason'' mode found in the NS junction 
\cite{elsewhere}. The lifetime of this mode is due to the coupling with the 
relaxation mode in the N-layer

\begin{equation}
\omega_2'=Q\left(
\frac{\pi\Delta D_s(\nu_s+\nu_m)}{\hbar \nu_m}\right)^{1/2};
\; \; 
\omega_2''= {\pi\over{2}} \left({{G_s}\over{G_m}}\right)
{\Delta\over{\hbar}}.
\label{oi}
\label{eq:6}
\end{equation}

\noindent
Equations (\ref{oi}) and (\ref{fm}) are valid for $\omega_i' > \omega_i''$. 
For the phason mode this condition is satisfied already at small frequencies 
$\hbar\omega \simeq \hbar\omega_2' \simeq \Delta (G_{s}/G_{m}) \ll \Delta$. 
The condition for the odd mode is weaker, given the smallness of the asymmetry
parameter $(\hbar\nu_m \delta D)^2 \ll G_s G_m$.

Now let us consider what happens, when a {\em dc} - current is driven in the 
normal layer. The average currents in the metal are accompanied by the 
fluctuations known as the shot noise. Since the currents in the metal are 
coupled to those in the S-layers, it is natural to expect that in the 
superconductors the fluctuating currents appear as well, and consequently, the
phasons are generated. 

To include these fluctuations in our description of the S-N-S sandwich, we add 
Langevin sources 
$\delta\vec{j}_l$
to the current in the normal metal. Equation (\ref{curm}) takes the form

\begin{equation}
\vec{j}_m
= - D_m
{\bf\vec{\nabla}}
\left( e^2 \nu_m \varphi + e n_m\right)
+ \delta 
\vec{j}_l.
\label{cml}
\label{eq:7}
\end{equation}

\noindent
The Gaussian fluctuations 
$\delta\vec{j}_l$
are described by their correlator. Out of equilibrium, when the energy 
relaxation is negligible $\tau_\epsilon \to \infty$, the electronic 
distribution function $f(\epsilon)$ in the normal metal is the two-step 
function 

\begin{equation}
f_{ne}(\epsilon) = -{1\over{2}}\left [ 
\eta\left (\epsilon + {{eU}\over{2}}\right) + 
\eta\left (\epsilon - {{eU}\over{2}}\right) \right ],
\label{dne}
\end{equation}

\noindent
where $\eta(x)$ is the Heaviside function. In that case the non-equilibrium 
part of the correlator of the fluctuations can be written as \cite{shot}

\begin{equation}
\langle \delta j_l^\alpha \delta j_l^\beta \rangle_{\omega,Q} = {1\over{2}}
\delta_{\alpha\beta}e^2D_m\nu_m (eU -\hbar|\omega|)\eta(eU-\hbar|\omega|).
\label{eq:8}
\label{cc}
\end{equation}

The difference of superconducting phases of the S-layers 
$\theta = \theta_1 - \theta_2$ in the presence of the current fluctuations 
$\delta\vec{j}_l$
can be determined from the system of Eqs.~(\ref{leq}), and (\ref{cml}) in the 
first order in asymmetry 

\begin{equation}
\delta\theta = {1\over{2e\hbar}} {{\delta D}\over{\nu_m D_m}}
{{\pi\Delta\omega \; \; \delta
\vec{j}_l
\cdot
\vec{Q}}\over{(\omega^2-\omega_1^2(Q))(\omega^2-\omega_2^2(Q))}}.
\label{theta}
\label{eq:9}
\end{equation}

\noindent
Therefore, the correlator of the phase fluctuations has two well pronounced 
poles corresponding to the two collective modes in the S-N-S sandwich and is 
proportional to the applied voltage $U$.

To find the effect of the phase fluctuations on the Josephson current, we
need the phase fluctuations in a single point. With the help of 
Eqs.~(\ref{eq:8}) and (\ref{eq:9}), we find in the lowest order order in
asymmetry

\begin{equation}
\langle \delta\theta^2 \rangle_{\omega} =
\int \frac{d^2Q}{\left(2\pi\right)^2} 
\langle \delta\theta^2 \rangle_{\omega, Q}
= {\frac{1}{G_s\Delta}}
\frac{eU} {\hbar|\omega|} \eta(eU-\hbar|\omega|),
\label{pco}
\label{eq:10}
\end{equation}

\noindent
at ${\hbar|\omega|} <eU$ and $\langle \delta\theta^2 \rangle_{\omega}= 0$ 
at ${\hbar|\omega|} > eU$. Equation (\ref{pco}) is valid, provided 
$\hbar\omega \gg G_s \Delta /G_m $.

Note that in Eq.~(\ref{pco}) the asymmetry parameter $\delta D$  
have disappeared and therefore only the subleading terms depend on the 
asymmetry parameter $\delta D$. The leading term Eq.~(\ref{pco}) is the 
contribution of the odd mode Eq.~(\ref{fm}) to the integral in Eq.~(\ref{pco}).
This term dominates because when both modes Eq.~(\ref{fm}) and Eq.~(\ref{oi})
are well defined, the lifetime of the odd mode is always longer than the
lifetime of the even phason mode (provided the asymmetry parameter 
$\delta D \ll D_m, D_s$).
However, the asymmetry parameter can not be arbitrary small,
as the lifetime of the odd mode should be less than the escape time (which is
the time it takes to remove a phason from the system and thus is the 
characteristic relaxation time). As we consider the S-N-S sandwich of the
infinite size, the escape time is practically set to infinity, which is the
reason the asymmetry disappears from the odd mode contribution Eq.~(\ref{pco}).
In this case, the correlator Eq.~(\ref{pco}) is independent of the 
particular choice of asymmetry. The difference
between various asymmetry realizations becomes important if one considers
a situation when the odd mode is still ballistic, while the even phason mode
is already damped. We will not discuss this situation in this Letter.

We have found that in the presence of the normal layer the phase fluctuations 
in the superconductor are large due to the large number 
($\sim eU/\hbar\omega$) of the phasons. 
Now we are interested in the effect of these fluctuations on the Josephson 
current. In equilibrium it is determined by the difference of time-independent 
phases of the superconducting order parameter in two S-layers 
$J(\theta)\propto\sin(\theta^{(0)})$. In the non-equilibrium situation one has
to average over the phase fluctuations so the Josephson current becomes 
modified by a phase factor.

The supercurrent flow in Josephson junctions was analyzed by many authors
\cite{been,zaik}. Equation (\ref{jc}) expresses the Josephson current density
as a sum of contributions of individual electronic states weighted with
their distribution function. The exact form of these contributions 
$j(\epsilon, \theta)$ depends on the boundary conditions at the S-N
interface \cite{kupr}. In the case of diffusive junction the supercurrent
density in the lowest non-vanishing order of transparency can be written 
as \cite{asla}

\begin{mathletters}
\begin{eqnarray}
&&
J(\theta) = {{2e}\over{\hbar}} E_J^{(0)} L_f \sin(\theta) 
\langle e^{i\delta\theta (0)}\rangle_\theta ,
\label{jcd}\\
&&
\nonumber\\
&&
L_f = Re \int {{d\epsilon}\over{\sqrt{-i\epsilon E_T}}}
{{f(\epsilon)}\over{\sinh\sqrt{-\displaystyle{{i\epsilon}\over{E_T}}}}},
\label{lg}
\end{eqnarray}
\end{mathletters}

\noindent
where the transverse
Thouless energy is $E_T = \hbar D_n/d^2$ ($d$ is the width of the normal 
layer) and the overall scale is given by the bare Josephson energy
$E_J^{(0)} \simeq G_1 G_2 /\nu_m$, where $G_1$ and $G_2$ are the tunneling 
conductances (in units of $e^2/\hbar$) per 
unit area of the two N-S interfaces in the junction.

Strictly speaking, Eq.~(\ref{jcd}) is exact only for homogeneous in space
phase fluctuation. The effect of the
inhomogeneity can be estimated as $\hbar D_s Q^2 /\Delta$, while the main 
contribution comes from the odd phason mode with 
$D_s Q^2 \sim \hbar\omega^2/\Delta$, and so the correction is of the order 
$\hbar^2\omega^2/\Delta^2$ and is small since we consider frequencies 
$\hbar\omega \le eU \ll \Delta$. 

In equilibrium $f(\epsilon)$ is given by the Fermi distribution. There are
no divergent phase fluctuations, so the factor 
$\langle e^{i\delta\theta (0)}\rangle_\theta$ in Eq.~(\ref{jcd}) is just
a number which can be incorporated in the definition of $E_J^{(0)}$.
At low temperatures the dominant contribution to the integral in 
Eq.~(\ref{lg}) is due to small frequencies. At $E_T \gg T$
the critical current is given by \cite{asla}

\begin{equation}
J_c^{(0)}= {{2e}\over{\hbar}} E_J^{(0)} \ln \left({{E_T}\over{T}}\right).
\label{crit-eq}
\end{equation}

In the non-equilibrium situation the critical current~(\ref{crit-eq}) is 
modified by two effects. First, the distribution function deviates from the
Fermi distribution. At low temperatures it is given by the two-step 
function~(\ref{dne}).
The applied voltage $eU$ sets the lower limit for the integration in 
Eq.~(\ref{lg}) and for the real part of the integral we obtain

\begin{eqnarray}
L_f = 
- \ln \left ( {{\tanh^2\sqrt{{eU}\over{8E_T}} + \tan^2\sqrt{{eU}\over{8E_T}}}
\over{1+\tanh^2\sqrt{{eU}\over{8E_T}}\tan^2\sqrt{{eU}\over{8E_T}}}} \right ).
\label{log}
\end{eqnarray}

\noindent
Increasing the applied voltage changes the sign of the logarithm in 
Eq.~(\ref{log}), see dashed line on Fig.~\ref{fig:1}, corresponding 
to  the $\pi$-junction.

The second effect is the appearance of the fluctuation phase factor in 
Eq.~(\ref{jcd}). To evaluate the average over $\theta$ we use the phase 
correlator Eq.~(\ref{pco}), Fourier transformed to the time domain 

\begin{equation}
\langle e^{i\delta\theta (0)}\rangle_\theta = 
\exp\left({\displaystyle{-\int{{d\omega}\over{2\pi}}\langle \delta\theta^2 
\rangle_{\omega}}}\right).
\label{av}
\end{equation}

\noindent
The frequency integration in Eq.~(\ref{av}) diverges logarithmically 
$\langle \delta\theta^2 \rangle \sim \ln (eU/\epsilon_*)$.
At large frequencies it is cut off at $\hbar\omega =eU$, because 
at $\omega > eU$ there are no classical fluctuations, see Eq.~(\ref{pco}).

The infrared divergency in Eq.~(\ref{av}) is due to the fact that we have 
neglected the Josephson current in the continuity equation (\ref{nep}). When 
taken into account, it leads to the opening of the gap in the spectrum of 
the collective modes, $\epsilon_* = \sqrt{E_J/\nu_m}$ (where $E_J$ is the
Josephson energy per unit area). This gap, 
provides the infrared cut off in Eq.~(\ref{av}). It is essential that in the 
non equilibrium situation, the Josephson energy decreases with the increase
of the bias voltage $U$ and vanishes at the critical point. The actual value
of $E_J$ at a given $U$ should be determined from the self-consistency
equation [obtained by substitution of Eq.~(\ref{av}) into Eq.~(\ref{jcd})]

\begin{equation}
E_J = E_J^{(0)} L_f
\exp\left[- {{eU}\over{\pi\Delta G_s}}\ln 
\left({eU}\sqrt{{\nu_m}\over{E_J}}\right)\right].
\label{sce}
\end{equation}

\noindent
Solving Eq.~(\ref{sce}) we obtain
 
\begin{eqnarray}
E_J (U) = E_J^{(0)} L_f^{1\over{1-\alpha}}
\left( {{{E_J^{(0)}}\over{(eU)^2} \nu_m }}
\right)^{{\alpha}\over{1-\alpha}}.
\label{res}
\end{eqnarray}

\noindent
where $\alpha = eU/2\pi\Delta G_s$.
The expression Eq.~(\ref{res}) is valid when $eU > \sqrt{E_J/\nu_m}$. In 
this case the resulting $E_J$ becomes suppressed by the non-equilibrium 
(compared to its mean field value $E_J^{(0)} L_f$).

We now can write down the non equilibrium critical current as

\begin{equation}
J_c = {{2e}\over{\hbar}} E_J(U),
\end{equation}

\noindent
where the renormalized Josephson energy is given by Eq.~(\ref{res}). The 
dependence of the critical current on the bias 
voltage is illustrated on Fig.~\ref{fig:1}.

{
\narrowtext
\begin{figure}[ht]
\vspace{0.2 cm}
\epsfxsize=6.3cm
\centerline{\epsfbox{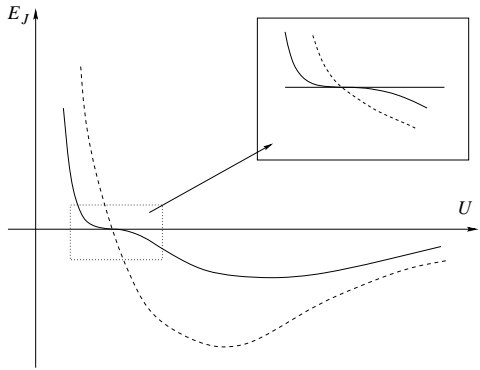}}
\vspace{0.2cm}
\caption{Plot of the Josephson energy Eq.~(\ref{res}) as a function of
the applied voltage $U$. The dashed line is the voltage-dependent logarithm
Eq.~(\ref{log}), which represents the critical current without taking into 
account the phase fluctuations. }
\label{fig:1}
\end{figure}
}

The expression for the Josephson energy Eq.~(\ref{res}) is the main
qualitative result of the paper. It shows the effects of non-equilibrium are
not limited to the change of the electronic distribution function. In addition,
one has to take into account the fluctuations of the superconducting phases in
both S-layers of the S-N-S junction. The phase fluctuations result in two 
observable effects. First, the Josephson energy at the critical point is not 
described by the mean field power law, but exhibits the non-analytic behavior
Eq.~(\ref{res}), illustrated by the inset on Fig.~\ref{fig:1}. Second, when 
the bias exceeds the critical value (so that the critical current becomes
negative) the Josephson energy is further suppressed relative to the mean
 field value.

We should warn the reader that Eqs.~(\ref{log}), (\ref{res}) 
were obtained for the simplest model of the S-N-S junction, 
namely the 2D sandwich. The spectrum of collective modes is sensitive to the 
geometry of the system, therefore, our results are  not expected to describe 
the experimental data ({\it e.g.} of Ref.~\onlinecite{exp3}) in detail. 
However, we have presented a strong evidence that the acoustic collective
modes, which are present in the junction in the non-equilibrium, can be 
observed by measuring the suppression of the Josephson energy. 

In conclusion, we showed that the non-equilibrium fluctuations of the 
superconducting phases in the S-N-S junction lead to the non-analytic behavior
of the Josephson energy at the critical point and to its suppression in the
region of the negative critical current, providing a possibility to observe 
the acoustic collective modes (phasons).

We acknowledge helpful conversations with L.I. Glazman.
I.A. is A.P. Sloan and Packard research fellow.

\end{multicols}


\begin{references}
\bibitem{exp1} H. Pothier, S. Gueron, N. O. Birge, D. Esteve, and M. H. 
Devoret, Phys. Rev. Lett. {\bf 79}, 3490 (1997).
\bibitem{exp2} A.F. Morpurgo, T.M. Klapwijk, and B.J. van Wees, Appl. Phys. 
Lett. {\bf 72}, 966 (1998).
\bibitem{been} See, e.g. C.W.J. Beenakker, Phys. Rev. Lett. {\bf 67}, 3836
(1991) and references therein; a review on transport in hybrid 
superconducting structures is C.J. Lambert and R. Raimondi, J. Phys.:
Condens. Matter {\bf 10}, 901 (1998).
\bibitem{exp3} J.J.A. Baselmans, A.F. Morpurgo,  B.J. van Wees, and 
T.M. Klapwijk, Nature {\bf 397}, 43 (1999).
\bibitem{zaik} L.N. Bulaevskii, V.V. Kuzii, and A.A. Sobyanin, Solid St.
Comm. {\bf 25}, 1053 (1977);
A.F. Volkov, Phys. Rev. Lett. {\bf 74}, 4730 (1995);
A.F. Volkov and H. Takayanagi, Phys. Rev. B {\bf 56}, 11184 (1997);
F. K. Wilhelm, G. Sch\"on, and A.D. Zaikin, Phys. Rev. 
Lett. {\bf 81}, 1682 (1998);
P. Samuelsson, J. Lantz, V.S. Shumeiko, and G. Wendin, cond-mat/9904276.
\bibitem{elsewhere} B.N. Narozhny, I.L. Aleiner, and
B.L. Altshuler, cond-mat/9903239.
\bibitem{Tinkham} M. Tinkham, 
{\em Introduction to Superconductivity} (McGraw - Hill, New York, 1996).
\bibitem{schm}A. Schmid and G. Sch\"on, Phys. Rev. Lett. {\bf 34}, 941 (1975);
R.V. Carlson and A.M. Goldman, Phys. Rev. Lett. {\bf 34}, 11 (1975);
S.N. Artemenko and A.F. Volkov, Zh. Eksp. Teor. Fiz. {\bf 69}, 1764 (1975)
[Sov. Phys. JETP {\bf 28}, 896 (1976)].
\bibitem{shot}Sh.M. Kogan and A.Ya. Shul'man,
Zh. Eksp. Teor. Fiz. {\bf 56}, 862 (1969) [Sov. Phys. JETP {\bf 29}
467 (1969)]; S.V. Gantzevich, V.L. Gurevich, and
R. Katilus {\it ibid} {\bf 57}, 503 (1969).
\bibitem{kupr} A.V Zaitsev, Zh. Eksp. Teor. Fiz. {\bf 86}, 1742 (1984) 
[Sov. Phys. JETP {\bf 59}, 1015 (1984)]; M.Yu. Kuprianov and V.F. Lukichev,
Zh. Eksp. Teor. Fiz. {\bf 94}, 139 (1988) [Sov. Phys. JETP {\bf 67}, 1163
(1988).
\bibitem{asla} L.G. Aslamazov, A.I. Larkin, and Yu.N. Ovchinnikov,
Zh. Eksp. Teor. Fiz. {\bf 55}, 323 (1968) [Sov. Phys. JETP {\bf 28}, 
171 (1969).
\end{references}
\end{document}